\documentclass[aps,twocolumn,showpacs]{revtex4}
\usepackage{graphicx}
\begin{document}
\title{Stability of shortest paths in complex networks with
random edge weights}
\author{Jae Dong Noh}
\affiliation{Theoretische Physik, Universit\"at des Saarlandes, 66041
Saarbr\"ucken, Germany}
\author{Heiko Rieger}
\affiliation{Theoretische Physik, Universit\"at des Saarlandes, 66041
Saarbr\"ucken, Germany}
\date{\today}

\begin{abstract}
We study shortest paths and spanning trees of complex networks with random
edge weights. Edges which do not belong to
the spanning tree are inactive in a transport process within the network.
The introduction of quenched disorder modifies the spanning tree such that
some edges are activated and the network diameter is increased. 
With analytic random-walk mappings and numerical analysis, 
we find that the spanning tree
is unstable to the introduction of disorder and displays a 
phase-transition-like behavior at zero disorder strength
$\varepsilon=0$. In the infinite network-size limit~($N\rightarrow \infty$), 
we obtain a continuous transition with the density of activated edges
$\Phi$ growing like $\Phi \sim \varepsilon^1$ and with the 
diameter-expansion coefficient $\Upsilon$ growing like 
$\Upsilon\sim \varepsilon^2$ in the regular network, 
and first-order transitions with discontinuous jumps
in $\Phi$ and $\Upsilon$ at $\epsilon=0$ for the small-world~(SW) network
and the Barab\'asi-Albert scale-free~(SF) network.
The asymptotic scaling behavior sets in when $N\gg N_c$, where the crossover
size scales as $N_c\sim \varepsilon^{-2}$ for the regular network,
$N_c \sim \exp[\alpha \varepsilon^{-2}]$ for the SW
network, and $N_c \sim \exp[\alpha |\ln \varepsilon| \varepsilon^{-2}]$
for the SF network.
In a transient regime with $N\ll N_c$, there is an infinite-order transition 
with $\Phi\sim \Upsilon \sim \exp[-\alpha / 
(\varepsilon^2 \ln N)]$ for the SW network and $\sim \exp[ -\alpha / 
(\varepsilon^2 \ln N/\ln\ln N)]$ for the SF network.
It shows that the transport pattern is practically most stable in the SF
network.
\end{abstract}
\maketitle

\section{Introduction}\label{sec:intro}
A network is a new paradigm to study complex systems in many 
disciplines in science~\cite{Strogatz01,AB02}. 
A complex system
consists of a large number of interacting units, and the 
nature of the interaction determines equilibrium and
dynamical properties of the system. Frequently the simplifying assumption 
is made that the units are arranged to form a simple pattern like a regular 
lattice or to interact with all others as in a mean-field theory.
Recent studies, however, have revealed that the structure of
complex systems is much richer~\cite{Strogatz01,AB02,Watts98}. 
In general this structure is captured by a network which consists of vertices 
representing the units and edges connecting interacting vertex pairs.

Complex networks exhibit so-called small-world phenomena: vertices are
highly clustered and the average separation between vertices grows slowly
with the total number of vertices. 
Watts and Strogatz~\cite{Watts98} introduced a small-world~(SW) network
as a model for these phenomena. 
It is obtained from a regular lattice with edges randomly rewired with 
probability $p_r$. 
Later it was found that some complex networks have a power-law distribution 
$P_{deg.}(z) \sim z^{-\gamma}$ of the degree $z$. 
The degree of a vertex is the number edges incident upon it.
The class of networks with a power-law degree distribution is called
the class of scale-free~(SF) networks and is found in many areas 
including physics, 
computer science, biology, sociology, etc~(we refer readers to 
Ref.~\cite{AB02} for examples). 
The Barab\'asi-Albert model~\cite{BA99} generate a SF network~($\gamma=3$)
{\em growing} via a {\em preferential attachment} rule~\cite{BAsolution}. 
Initially one starts with $Z_0$ vertices and 
introduces a new vertex at each step~[growing]. It is then attached to 
$Z$ existing vertices, which are selected with probability 
{\em linearly} proportional to their degree~[preferential
attachment]. 

The discovery of the new classes of networks triggers extensive research.
Order-disorder phase transitions~\cite{Dorogovtsev,Igloi,KimBJ} and 
nonequilibrium phase transitions~\cite{Pastor-Satorras}
have been studied. Interestingly, the critical behavior is described well
by mean-field theories and strongly depends on the degree
distribution~\cite{Igloi}.
Stability of complex networks has also been studied against a strong
disorder such as a vertex dilution~\cite{Albert00,Cohen00,Callaway00}. 
When the fraction of diluted vertices increases, a network may 
disintegrate into finite clusters undergoing a percolation-type transition.
In this paper, we study the effect of weak disorder on the transport
properties of networks.

The {\em shortest path} plays an important role for the transport within a
network~\cite{Goh01,Szabo02}. A {\em path} denotes a sequence of vertices, 
successive pairs of which are connected via edges. 
In general there exist many paths connecting two given vertices. 
The shortest path is the one with minimum path length among all the paths.
The minimum path length is called a {\em separation} between
the two vertices. Suppose one needs to send, e.g., a data packet from one 
computer to the other through the Internet. The shortest path provides
an optimal path way since one would achieve a fastest transfer
and save system resources. The shortest path is also important in studying
an internal structure of a network~\cite{Newman01}. 
The separation can be used as a measure of intimacy between vertices.
The number of shortest paths that pass through a vertex is called the
``betweenness" or ``load"~\cite{Newman01,Goh01,Szabo02,Goh02}. 
It reflects the importance of a vertex in mutual relationship or in transport.
The load follows universal power-law distributions in
scale-free networks~\cite{Goh01,Szabo02,Goh02}.

\begin{figure}
\includegraphics[width=\columnwidth]{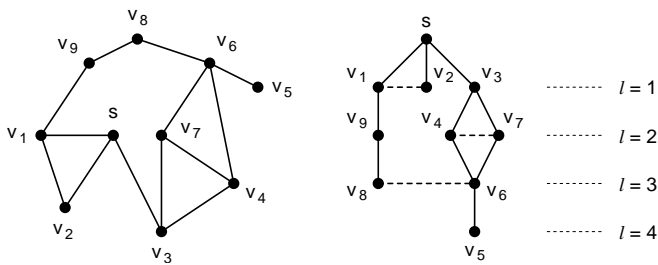}
\caption{A network (left) and its spanning tree (right) ${\bf T}_s$.
Vertices in ${\bf T}_s$ are arranged in the hierarchical order of the
separation $l$.
The vertex $v_6$, and hence $v_5$, has twofold degenerate shortest paths, 
so ${\bf T}_s$ contains a loop. Dashed lines represent edges which do not
belong to ${\bf T}_s$.}
\label{fig:sp}
\end{figure}

Consider shortest paths from a vertex
$s$, called the {\it source}, to all other vertices in a network. 
In an un-weighted network the length of a path is just the number of edges
that it contains and shortest paths can simply be found using the 
``breadth-first search algorithm"~\cite{Heiko}. 
A sub-network consisting of all the shortest paths from $s$ 
is the {\em spanning tree} ${\bf T}_s$, which characterizes the optimal 
transport pattern.
Figure~\ref{fig:sp} shows an example of a network and its spanning tree.
It is convenient to represent the spanning tree by a diagram in which 
vertices are arranged hierarchically in the ascending order of their 
separation from the source. 
Then, the shortest path to each vertex is given by a 
directed path on ${\bf T}_s$.
In general, the spanning tree does not have a tree structure. 
If there are degenerate shortest paths, ${\bf T}_s$ contains a loop.
Some edges do not belong to ${\bf T}_s$. They do not contribute 
to any flow from/to $s$.

It has been assumed that all edges are equivalent having the same cost.
However a real network would be described better with weighted edges.
A weight of an edge may represent an access cost, a physical length,
or an intimacy between vertices~\cite{Newman01,Yook01}.
For example, edges between scientists in scientific collaboration networks
may have weights which depends on the number of coauthored 
papers~\cite{Newman01}. 
In a weighted network, the path with the minimum number of edges is not
necessarily an optimal one. 
In this work, we study disordered networks with randomly-weighted
edges and investigate the effect of the disorder on the transport pattern.
The regular network, the SW network~\cite{Watts98}, and the 
Barab\'asi-Albert SF network~\cite{BA99} are considered.
This paper is organized as follows:
In Sec.~\ref{sec:sp&st}, we define the shortest path and the spanning tree
in the disordered network. 
The disorder modifies the shape of the spanning tree. The response is described 
for the regular and SW networks in Sec.~\ref{sec:sw} 
and for the SF network in Sec.~\ref{sec:BA}. We conclude 
in Sec.~\ref{sec:conclusion}.

\section{Shortest paths and the spanning tree of disordered networks}
\label{sec:sp&st}
Consider a disordered undirected network. An edge $e$ between two vertices 
$u$ and $v$ will be denoted as $e=[u;v]=[v;u]$.
To each edge $e$ 
a non-negative weight $c(e)$ is assigned which is called 
the {\em edge cost} of $e$.
Here we neglect all system-dependent details and assume that 
\begin{equation}\label{edge-cost}
c(e) = 1 + \eta(e),
\end{equation}
where $\eta(e)$'s are random variables distributed independently with 
distribution ${\cal F}(\eta)$~($\eta(e)>-1$). 

In a disordered network, a {\em minimum-cost
path} plays the role of the shortest path in a pure~($\eta=0$) 
network. For given vertices $u$ and $v$, the minimum-cost path  
is given by the one with minimum {\em path cost}. 
The path cost is defined as the sum of all edge costs in the path.
Without disorder~($\eta(e)=0$ for all $e$), the path cost is equivalent to
the path length and the minimum-cost path is the same as the
shortest path.  
Hereafter, the minimum-cost path will be called the shortest path,
and the minimum cost is denoted as a 
{\em distance}. The path length of the shortest path is called
the {\em separation} of the two vertices connected by it. 

The spanning tree ${\bf T}_s$ of a disordered network 
can be found using the ``Dijkstra algorithm"~\cite{Heiko}: 
Divide all vertices into two sets $S$ and its complement $\bar{S}$. 
Initially $S=\{s\}$ and the source is assigned to a distance label $d(s)=0$
and a separation label $l(s)=0$. 
At each iteration, one selects an optimal edge $e^\star=[u^\star;v^\star]$ 
that has a minimum value of $d(u)+c([u;v])$ among all edges $e=[u;v]$ with 
$u\in S$ and $v\in\bar{S}$. Then, the vertex $v^\star$ gets
the labels $d(v^\star) = d(u^\star)+c([u^\star;v^\star])$ and
$l(v^\star) = l(u^\star)+1$ and a predecessor label $pred(v^\star)=u^\star$, 
and is shifted from $\bar{S}$ to $S$.
The iteration terminates when the set $\bar{S}$ is empty. The shortest path
to each vertex is then found by tracing the predecessor iteratively back to
$s$. The distance and separation from $s$ to each vertex $v$ 
are given by $d(v)$ and $l(v)$, respectively. 
The average separation $D_s = \frac{1}{N-1}\sum_{v\neq s} l(v)$ will be 
called a {\em diameter}. 

In a homogeneous~(e.g. regular) network and a weakly disordered
network~(e.g. SW network), all vertices are equivalent after an
average over all disorder realizations. 
The diameter is independent of the source, $D_s = D$.
In such cases we select the source $s$ arbitrarily. 
On the other hand, the SF network has a highly inhomogeneous
structure. For the SF network, we select the {\em hub} which has the largest
degree as the source $s$ since it plays the most important role in the
transport~\cite{Goh01}.

The spanning tree ${\bf T}_s$ of a disordered network will be different
from ${\bf T}_s^0$ of the same network without disorder. 
If the disorder has a continuous distribution, with probability one all
shortest paths are uniquely determined.
Therefore, ${\bf T}_s$ has a 
tree structure, whereas ${\bf T}_s^0$ may have loops. 
Without loops, ${\bf T}_s$ consists of only $(N-1)$ edges.
Moreover, a vertex may have a shortest path that cannot be found 
in ${\bf T}_s^0$.  
For example, a path $(s,v_1,v_9,v_8,v_6)$ in a network in
Fig.~\ref{fig:sp} may be a shortest path to $v_6$, so that ${\bf T}_s$ 
includes the edge $[v_8;v_6]$ which does not belong to ${\bf T}_s^0$. 
Such an edge of ${\bf T}_s$ that does not belong to ${\bf T}_s^0$ will be 
called an {\em activated} edge. The disorder activates it to play a role in
the transport.
The activated edge results in a drastic change in the shape of
the spanning tree and increases the network diameter. 
We quantify the change by the density of disorder-induced activated edges 
$\Phi$, which is given by the number
of activated edges in ${\bf T}_s$ divided by $(N-1)$, and the
diameter-expansion coefficient $\Upsilon = (D_s-D_s^0)/D_s^0$ with 
$D_s$~($D_s^0$) the diameter with~(without) disorder.
The activated edge emerges as a result of competition between all paths 
connecting a vertex to the source. So networks with different structures
respond differently. 
In the next section we will study the evolution of the spanning trees of 
the regular network and the small-world network~\cite{Watts98}.

\section{Small-world network}\label{sec:sw}
Consider a regular network consisting of $N$ vertices on a
one-dimensional ring, each of which is connected up to $Z$th nearest
neighbors with undirected edges. The small-world~(SW) network is 
obtained by rewiring each edge with probability $p_r$~(see 
Ref.~\cite{Watts98} for a detailed procedure). 
Except for extreme cases with $p_r=0$~(regular network) and 
$p_r=1$~(random network), the SW network
displays the small-world phenomena~\cite{Watts98}.
We introduce a quenched disorder 
in edge costs as in Eq.(\ref{edge-cost}) with the disorder distribution 
\begin{equation}\label{eta-distri}
{\cal F}(\eta) = \left\{ 
\begin{array}{cll}
 \frac{1}{2\varepsilon} & \quad &\mbox{for\ } 
 -\varepsilon\leq \eta< \varepsilon \\ [2mm]
 0 &\quad & \mbox{otherwise.}
\end{array}\right. 
\end{equation}
Disorder strength is controlled by the parameter $\varepsilon$~($<1$).
First we focus on the regular network~($p_r=0$) with $Z=2$, which gives
us a lot of insights. 

\subsection{Regular network ($p_r=0$) with $Z=2$}
\begin{figure}
\includegraphics[width=\columnwidth]{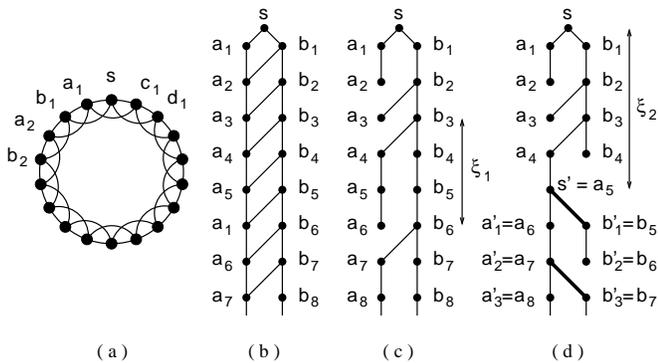}
\caption{A regular network with $Z=2$ in (a) and its spanning trees
without disorder in (b) and with disorder in (c) and (d). (c) shows
a spanning tree of type B and (d) shows a spanning tree with both
segments of type A and type B. Vertices in (b), (c), and (d) 
are arranged in the hierarchical order in the separation from the source $s$.
Activated edges are represent by thick lines in (d).}
\label{fig:regular}
\end{figure}
A regular network with $Z=2$ and $N=18$ is shown in
Fig.~\ref{fig:regular}(a). We consider the shortest path from 
the source $s$. Without disorder~($\varepsilon=0$), the shortest path to 
$b_n$ is unique. On the other hand, there are $n$
degenerate shortest paths to $a_n$. So the spanning tree ${\bf T}_s^0$ 
has a ladder shape with diagonal rungs from
$b_n$ to $a_{n+1}$ as shown in Fig.~\ref{fig:regular}(b)~\cite{comment}.
All edges $\{[a_n;b_n]\}$ are missing in ${\bf T}_s^0$.

With infinitesimal disorder, all loops break up in the spanning tree. So
either $[a_{n-1};a_n]$ or $[b_{n-1};a_n]$ should be removed from ${\bf T}_s$.
Consequently, ${\bf T}_s$ has a tree shape with a single branch for $b_n$'s 
and with sub-branches for $a_n$'s, cf., Fig.~\ref{fig:regular}(c). 
This shape will be denoted as a type-B tree.
The branching points are determined from recursion relations for the distances
$d(a_n)$ and $d(b_n)$: 
\begin{eqnarray}
d(a_n) &=& \min\{ d(a_{n-1}) + c([a_{n-1};a_n]) ,  \nonumber \\
       & & \quad  d(b_{n-1}) + c([b_{n-1};a_n])\} \label{d_an}\\
d(b_n) & = & d(b_{n-1}) + c([b_{n-1};b_n]) \ . \label{d_bn}
\end{eqnarray}
If $d(a_{n-1}) + c([a_{n-1};a_n]) < d(b_{n-1}) + c([b_{n-1};a_n])$, the
predecessor of $a_n$ is $a_{n-1}$, otherwise the predecessor is
$b_{n-1}$.
The recursion relation~(\ref{d_an},\ref{d_bn}) holds if ${\bf T}_s$ has 
the type-B structure,
which is valid as long as 
\begin{equation}\label{constraint}
d(a_n) + c([a_n;b_n]) > d(b_n).
\end{equation}
When it is violated, the predecessor of $b_n$ is $a_n$, 
$[a_n;b_n]$ becomes activated, and ${\bf T}_s$ changes its shape.

We introduce now a random walk interpretation of the recursion relation.
Define $X_{A} (n) \equiv d(a_n) - n$ and $X_{B} 
\equiv d(b_n)-n$ and insert it into Eq.~(\ref{d_bn}). 
Using Eq.~(\ref{edge-cost}), 
one obtains $X_B(n) = X_B(n-1) + \eta([b_{n-1};b_n])$;
$X_B(n)$ can be interpreted as the coordinate of
a one-dimensional random walker~(walker $B$) after $n$ jumps. 
$X_A(n)$ is given by the minimum of
$X_A(n-1)+\eta([a_{n-1};a_n])$ and $X_B(n-1) + \eta([b_{n-1};a_n])$.
The first term also suggests that $X_A(n)$ can be interpreted as the coordinate 
of another random walker~(walker $A$) after $n$ jumps. But its motion
is constrained: After each jump, one has to compare the 
current position of $A$~[the first term] with the position of $B$~[the second 
term$\simeq X_B(n)+{\cal O}(\varepsilon)$], and take the minimum as $X_A(n)$.  
Therefore, one may assume a hard-core repulsion
between two walkers~(the interaction range fluctuates by an amount of 
${\cal O}(\varepsilon)$). 
The inequality in Eq.~(\ref{constraint}) can be rewritten as
$X_B(n) - X_A(n) < c([a_n;b_n]) = 1 + {\cal O}(\varepsilon)$,
which imposes a constraint on the random-walk motion.

It is convenient to introduce $X_R(n) \equiv X_B(n) - X_A(n)$. 
It can be interpreted as
the coordinate of a one-dimensional random walker $R$ in the presence of a
fluctuating {\em reflecting} wall at $X_R={\cal O}(\varepsilon)$ 
and a fluctuating {\em absorbing} wall at $X_R=1+{\cal O}(\varepsilon)$. 
At each time step, the walker $R$ 
performs a jump of size $\eta'$ obeying the distribution 
${\cal F}_R(\eta')\equiv \int d\eta_1\int d\eta_2 {\cal F}(\eta_1){\cal
F}(\eta_2) \delta(\eta'-\eta_1+\eta_2)$.
Hereafter the boundary walls are assumed to be fixed at $X_R=0$
and $X_R=1$. The fluctuations do not modify the scaling behavior
of $\Phi$ and $\Upsilon$ with a possible change in the coefficient of the
leading order term.

The random walker $R$ determines the shape of the spanning tree.
If the walker does not touch either wall at $X_{R}=0$ or $1$ at a 
moment $n$, then the predecessor of $a_n$~($b_n$) is $a_{n-1}$~($b_{n-1}$).
When it bounces at the reflecting wall in step $n$, 
$a_n$ has $b_{n-1}$ as its predecessor, and the
spanning tree has a new sub-branch, cf. 
$a_3$, $a_4$, and $a_7$ in Fig.~\ref{fig:regular}(c). 
If it collides with the absorbing wall 
at step $n_0$, the inequality (\ref{constraint}) is violated and
$b_{n_0}$ has $a_{n_0}$ as predecessor instead of $b_{n_0-1}$~(see
the spanning tree in Fig.~\ref{fig:regular}(d) which has $n_0=5$). 

The same random-walk mapping can be established after an edge
$[a_{n_0};b_{n_0}]$ is activated. If one interprets $s'=a_{n_0}$ 
as a new source 
and re-defines $a'_n = a_{n+n_0}$ and $b'_n = b_{n+n_0-1}$, 
the distances from $s'$ to $a'_n$ and $b'_n$ satisfy the same recursion
relation as in Eqs.~(\ref{d_an}) and (\ref{d_bn}) and
the same constraint as in Eq.~(\ref{constraint}) with $a~(b)$ replaced by
$b'~(a')$. Thus, the spanning tree consists of a single branch for $a'$'s
and sub-branches for $b'$'s. This shape will be denoted as a type A,
see Fig.~\ref{fig:regular}(d). 
The creation of sub-branches and the 
switch into type-A tree are described by the same random-walk mapping used 
for the type-B segment.

Combining the mappings for type-A and type-B segments,
the shape of the whole spanning tree can be determined by the
the random walker $R$ in the presence of 
two hard-core walls at $X_{R}=0$ and $1$.
Initially $X_{R}(0)=0$, and the wall at $X_{R}=0~(1)$ is
reflecting~(absorbing). 
When the absorbing wall is at $X_{R}=1~(0)$, 
the spanning tree has the type-B~(type-A) shape.
The sub-branch emerges when the random walker collides with the
reflecting wall. When it collides with the absorbing wall, the role of the
two walls are exchanged and the spanning tree switches its shape.
It is interesting to note
that the random walk with two types of boundary walls were used to find
the exact ground states of one-dimensional random-field Ising-spin 
chain~\cite{Schroder}.

The shape of ${\bf T}_s$ is characterized by two length scales $\xi_1$
and $\xi_2$, see Fig.~\ref{fig:regular}. The former characterizes the
length of the sub-branch, and the latter the
length of each type-A or type-B segment. They are given by the mean time 
scales between successive collisions with the reflecting wall and the
absorbing wall, respectively.
Then, $\xi_1$ can be approximated as the life time of the random walker $R$, 
being at the origin initially, in the presence of two absorbing walls 
at $X=0$ and $X=1$.
And $\xi_2$ is given by the mean life time of the random walker $R$, 
being at the origin initially, in the presence of two absorbing walls at 
$X=\pm 1$. Such time scales are calculated in time-continuum limit in
Appendix. 
Using the results in Eqs.~(\ref{tau1}) and (\ref{tau2}) with $a=\sigma$ 
and $\sigma^2\equiv \int d\eta' {\cal F}_R(\eta') \eta'^2
= \frac{2}{3}\varepsilon^2$, we obtain that
$\xi_1  \simeq  \sqrt{3/2}~\varepsilon^{-1}$ and 
$\xi_2  =  (3/2) \varepsilon^{-2}$.
Note that the ${\cal O}(\varepsilon)$ fluctuation of the walls does 
not change the scaling exponents, but may modify the coefficient of $\xi_1$.

\begin{figure}
\includegraphics[width=\columnwidth]{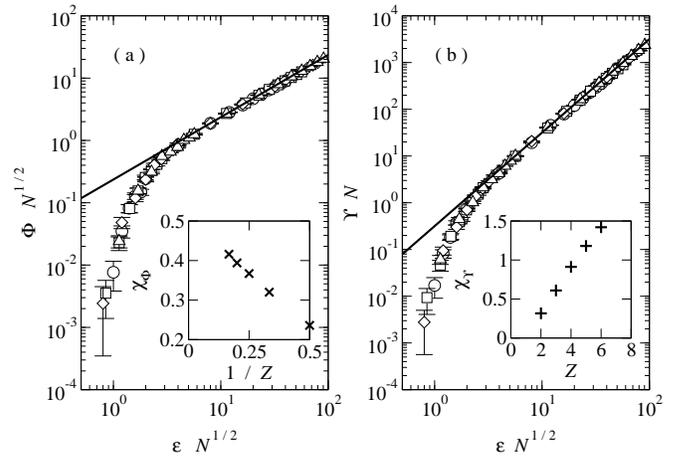}
\caption{$\Phi_{REG}$ in (a) and $\Upsilon_{REG}$ in (b) for the regular 
network with $Z=2$ and $N=10000~(\circ)$, 20000~($\fbox{}$), 
40000~($\diamond$), and 80000~($\bigtriangleup$).  
The straight lines represent $0.24 x$ in (a) and $0.32 x^2$ in (b).
Insets show $Z$ dependence of the coefficients 
$\chi_\Phi$~($\Phi_{REG}\simeq \chi_\Phi \varepsilon$) and 
$\chi_\Upsilon$~($\Upsilon_{REG} \simeq \chi_\Upsilon\varepsilon^2$).}
\label{fig:reg_num}
\end{figure}

One edge among $Z$ edges in a row is activated when a sub-branch appears
only in the type-A segment~(see Fig.~\ref{fig:regular}(d)). 
So, the activated-edge density is inversely proportional to $\xi_1$: 
\begin{equation}\label{phi2}
\Phi_{REG}(\varepsilon) = \frac{1}{Z}\frac{Z-1}{Z}\ \xi_1^{-1} \simeq 
\frac{1}{\sqrt{24}}\ \varepsilon ,
\end{equation}
where $Z=2$ and $(Z-1)/Z$ is the fraction of the type-A segments.
When the spanning tree changes its shape at a certain vertex~(cf. $a_5$ in
Fig.~\ref{fig:regular}(d)), its all descendent vertices 
increases their separation from $s$ by one. 
So, the diameter-expansion coefficient is
inversely proportional to $\xi_2$:
\begin{equation}\label{upsilon2}
\Upsilon_{REG}(\varepsilon) = \frac{1}{Z}\ \xi_2^{-1} = 
\frac{1}{3}\ \varepsilon^2 \ .
\end{equation}
The results are valid in the asymptotic limit where 
$N \gg \max[\xi_1,\xi_2] = \xi_2$, which suggests
the finite-size-scaling form 
$\Phi_{REG}(\varepsilon,N) = N^{-1/2} {\cal G}_\Phi (\varepsilon N^{1/2})$
and $\Upsilon_{REG}(\varepsilon,N) = 
N^{-1} {\cal G}_\Upsilon (\varepsilon N^{1/2})$.
The scaling functions behave as ${\cal G}_\Phi (x) \simeq \chi_\Phi x$ and 
${\cal G}_\Upsilon(x) \simeq \chi_\Upsilon x^2$ for $x\gg 1$.
The scaling behavior is confirmed numerically. 
We compute both quantities for the regular network with $Z=2$ and
$N=10000,\ldots,80000$, which were averaged over 200 samples. 
They are plotted in Fig.~\ref{fig:reg_num}, where data collapse very well. 
From a least-square fitting we obtained that 
$\chi_\Phi \simeq 0.24$ and $\chi_\Upsilon \simeq 0.32$, which is close
to the analytic results (\ref{phi2}) and (\ref{upsilon2}).

The extension to the regular networks with $Z>2$ is straightforward. 
Vertices are labelled as $(s,v_1^1,\ldots,v_1^Z$, $v_2^1,\ldots,v_2^Z,\ldots)$
starting from the source $s$.
Then, without disorder, the spanning tree ${\bf T}_s^0$ has a $Z$-leg ladder
structure with diagonal rungs. 
Each leg~($i=1,\ldots,Z$) consists of vertices $\{v_n^i\}$. 
The predecessors of a node $v^i_{n+1}$ are $v^j_n$ with $j=i,\ldots,Z$.
Edges $\{[v^i_n;v^j_n]\}$ with $i\neq j$
do not belong to ${\bf T}_s^0$. When the disorder turns on, there emerge
activated edges. Until one finds an activated edge, the distance to
each vertex from the source satisfies recursion relations
\begin{equation}\label{dZ}
d(v^i_n) = \min_{j=i,\ldots,Z} \{ d(v^j_{n-1}) + c([v^j_{n-1};v^i_n]) \} \ .
\end{equation}
The recursion relations are valid as long as 
\begin{equation}\label{constraintZ}
d(v^i_n) + c([v^i_n;v^j_n]) > d(v^j_n)
\end{equation}
for all $i\neq j$.
With the mapping $X_i(n) = d(v^i_n) - n$, one can 
interpret $X_i(n)$ as a coordinate of a random
walker $A_i$ after $n$ jumps~(each jump has the distribution ${\cal
F}(\eta)$). Then, Eq.~(\ref{dZ}) implies a hard-core interaction between
walkers, so $A_i$ cannot overtake $A_{j}$'s with
$j>i$. Effectively, it suffices to consider the hard-core
interaction only between $A_i$ and $A_{i+1}$. 
If $A_i$ and $A_{i+1}$ collide at step $n$, 
then the vertex $v^i_{n}$ takes $v^{i+1}_{n-1}$ as its predecessor. 
Otherwise, $v^{i}_{n-1}$ is the predecessor of $v^i_{n}$. The constraint
(\ref{constraintZ}) implies that the relative distance between all walkers 
should be less than $1+{\cal O}(\varepsilon)$, that is, $|X_1-X_Z| \lesssim 1$. 
An activated edge $[a^1_{n_0};a^Z_{n_0}]$ appears when this inequality 
is violated in the $(n_0+1)$th step. Then, we can use the same random-walk
mapping after a cyclic permutation $(A_1,\ldots,A_Z)\rightarrow
(A_2,\ldots,A_Z,A_1)$, which continues repeatedly.
As in the case with $Z=2$, the shape of the spanning tree is characterized
by the length scale $\xi_1$, the mean time scale for a collision between 
adjacent walkers, and $\xi_2$, the mean time scale for violating the
constraint $|X_1-X_Z| \lesssim 1$. 
The time scales are approximately equal to those for the two-random-walker
problem with a constraint $|X_1-X_2| \lesssim {2}/{Z}$. 
With this approximation, one gets
$\xi_1 \sim (Z\varepsilon)^{-1}$ and 
$\xi_2 \sim (Z\varepsilon)^{-2}$. Using $\Phi_{REG} =
\frac{Z-1}{Z^2 \xi_1}$ and $\Upsilon_{REG} = 
\frac{1}{Z\xi_2}$, we finally obtain
$\Phi_{REG} \simeq \chi_\Phi \varepsilon^{1}$ and $\Upsilon_{REG} \simeq
\chi_\Upsilon \varepsilon^{2}$ with $Z$-dependent coefficients
$\chi_\Phi \propto (1-\frac{1}{Z})$ and $\chi_\Upsilon \propto Z$.
The scaling exponents are universal for all $Z$. 
We determine the
coefficients of $\varepsilon$ and $\varepsilon^2$ for $\Phi_{REG}$ and 
$\Upsilon_{REG}$, respectively, numerically and plot them as a function of 
$Z$ in the insets of Fig.~\ref{fig:reg_num}.
One sees that $\chi_\Phi \propto (1-1/Z)$ and $\chi_\Upsilon \propto Z$,
as estimated above.

We conclude that quenched disorder is a relevant perturbation to the 
spanning tree of the regular network. Using the random-walk mapping, 
we have shown that a finite fraction of edges in the spanning tree are 
modified at nonzero disorder strength. 
This fraction is linearly proportional to the disorder strength, 
$\Phi_{REG} \sim \varepsilon$. We have also shown that the diameter-expansion
coefficient is proportional to the square of the disorder strength, 
$\Upsilon_{REG}\sim \varepsilon^2$.

\subsection{Small-world network}
\begin{figure}
\includegraphics[width=0.8\columnwidth]{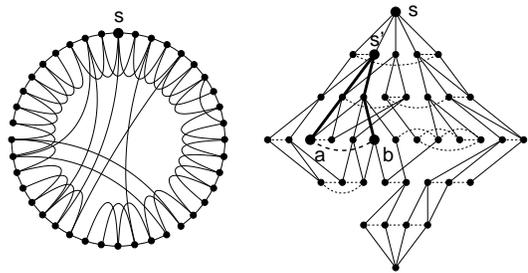}
\caption{A small-world network (left) with $N=40$, $Z=2$, and $p_r = 0.1$
and its spanning tree ${\bf T}_s^0$~(right) without disorder. 
The dashed lines indicate edges which do not belong to ${\bf T}_s^0$.}
\label{fig:sw}
\end{figure}

\begin{figure}
\includegraphics[width=\columnwidth]{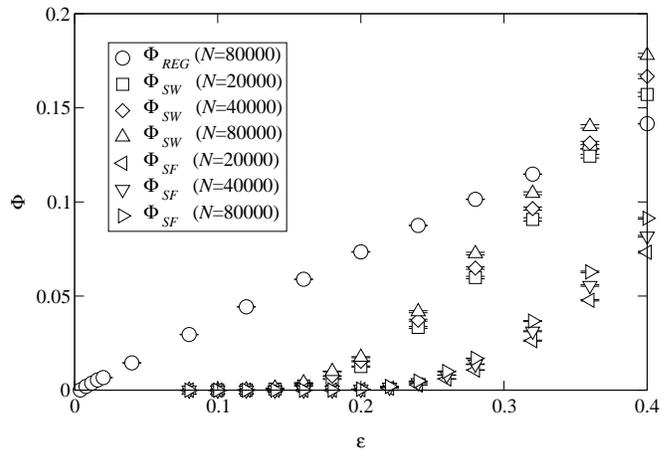}
\caption{Activated edge densities in the regular network $\Phi_{REG}$ 
with $Z=4$, the SW network $\Phi_{SW}$ with $Z=4$ and $p_r=0.2$,
and the SF network $\Phi_{SF}$ with $Z=Z_0=4$.}\label{fig:comp}
\end{figure}

Next we study the effect of the quenched disorder in the SW 
network. Figure~\ref{fig:sw} shows an example of a
SW network and its spanning tree ${\bf T}_s^0$ without
disorder~($\varepsilon=0$). The random rewiring of edges randomizes ${\bf
T}_s^0$, too,
which is not suited for an exact description. 
Therefore we study $\Phi_{SW}$ and $\Upsilon_{SW}$ first numerically.
We calculate $\Phi_{SW}$ in SW networks with $Z=4$ and $p_r=0.2$ 
and compare it with $\Phi_{REG}$ in Fig.~\ref{fig:comp}. 
$\Phi_{SW}$ shows two noticeable features:
(i) At small $\varepsilon$, $\Phi_{SW}$ appears to display a threshold behavior 
at a nonzero value of $\varepsilon$.
(ii) $\Phi_{SW}$ has a strong size dependence. 
$\Phi_{SW}$ does not approach an asymptotic saturation at any value of 
$\varepsilon$ and $N$ considered. 
The same feature is commonly observed at other values of $Z$ and $p_r$, 
and also for $\Upsilon_{SW}$.
In what follows we describe the response of the spanning tree with an
effective random walk process and we will show that the origin of the
apparent threshold behavior is the presence of a well-defined
$\varepsilon$-dependent crossover size in the network.

All edges that connect vertices at the same hierarchy level in ${\bf T}_s^0$, 
as represented by dashed lines in the example
in Fig.~\ref{fig:sw}, are candidates for an activated edge.
Focus on a pair of vertices $a$ and $b$ with $[a;b]\notin {\bf T}_s^0$ 
in Fig.~\ref{fig:sw}. They are descendants of a common ancestor $s'$.
The edge $[a;b]$ belongs to ${\bf T}_s$ if
a difference between costs of the two paths~(denoted by thick lines) 
from $s'$ to $a$ and to
$b$ is larger than the edge cost $c([a;b])=1+\eta([a;b])=
1+{\cal O}(\varepsilon)$, 
where ${\cal O}(\varepsilon)$ term can be neglected. 
The probability with which this happens will be denoted as 
$P_{act.}([a;b])$. 
Let $l(a,b)$ be the separation of $a$ and $b$ to their
common ancestor $s'$ in ${\bf T}_s^0$.
Using Eq.~(\ref{edge-cost}), the path costs are given by a sum of $l(a,b)$ 
independent random variables $\eta$'s (plus $l(a,b)$). 
So, with the common term $l(a,b)$ discarded, they can be interpreted as 
coordinates of two one-dimensional random walkers after $l(a,b)$ jumps.
Each jump follows the distribution ${\cal F}(\eta)$.
$P_{act.}([a;b])$ is then given by the probability that the distance 
between two random walkers is larger than $1$ after $l$ jumps. 
Or equivalently, it is given by the probability that a random walker 
deviates from a starting position by a distance larger than $1$ after $l$
jumps, where each jump follows a distribution ${\cal F}_R(\eta') =
\int d\eta_1 \int\eta_2 {\cal F}(\eta_1){\cal F}(\eta_2) \delta(\eta'-
\eta_1+\eta_2)$. The probability distribution of the random walker after $l$
steps is given by $P_{R.W.}(x,l) = (2\pi\sigma^2 l)^{-1/2}
e^{-x^2/(2\sigma^2 l)}$ with $\sigma^2 \equiv \int d\eta' \eta'^2 {\cal
F}_R(\eta') = 2\varepsilon^2/3$. Therefore, one obtains
\begin{equation}\label{p_ab}
P_{act.}([a;b]) = 
{\rm erfc}\left(\frac{\sqrt{3}}{2\varepsilon\sqrt{l(a,b)}}\right) \ ,
\end{equation}
where ${\rm erfc}(x)\equiv \frac{2}{\sqrt{\pi}} \int_x^\infty e^{-x^2}$ is 
the complementary error function.

Now we make a mean-field-type approximation that each edge $e\notin {\bf
T}_s^0$ may be an activated edge independently. 
That is only true for the edges $\{e_i=[a_i;b_i]\notin {\bf T}_s^0\}$ 
if the paths from $a_i$ and 
$b_i$ to their common ancestor do not overlap for different $e_i$'s.
Otherwise, two probabilities $P_{act.}(e)$ and $P_{act.}(e')$ 
are not independent. 
If $e$ is activated, it modifies the shape of
the spanning tree and hence $P_{act.}(e')$, and vice versa. The
approximation would give the correct scaling behavior if the spanning tree 
would be random and self-averaging.
In the mean-field scheme, the activated edge density is 
proportional to the probability in Eq.~(\ref{p_ab}) with $l(a,b)$ replaced
by the average value, i.e., the diameter of the network $D_s \sim \ln
N$~\cite{Newman00}: 
\begin{equation}\label{phi_sw}
\Phi_{SW}(N,\varepsilon) \simeq \alpha_1 \ 
{\rm erfc}\left(\frac{\alpha_2}{\varepsilon \sqrt{\ln N}}\right)
\end{equation}
with constants $\alpha_{1,2}$ being independent of $\varepsilon$ and $N$.

The scaling function behaves as ${\rm erfc}(x) \simeq 
1-\frac{2}{\sqrt{\pi}}x$ for small $x$. 
So, in the asymptotic limit where $N \gg N_c \simeq 
e^{\alpha_2^2 \varepsilon^{-2}}$, the activated edge density is a 
constant~[$\Phi_{SW} \simeq \alpha_1$] in the infinite~($N\rightarrow\infty$)
network for all values of $\varepsilon\neq 0$.
Therefore, the spanning tree of the SW network undergoes a discontinuous 
transition at $\varepsilon=\varepsilon_c=0$ in the asymptotic limit, which
is contrasted to the continuous transition in the regular network. 

The asymptotic behavior sets in when the network size is bigger than 
the crossover size $N_c\simeq e^{-\alpha_2^2\varepsilon^{-2}}$. 
It grows very fast as $\varepsilon$ goes to zero. 
For instance, when $\varepsilon=0.1$ with $\alpha_2\simeq 1$, 
only SW networks with $N \gg N_c \simeq  10^{43}$ are in the
asymptotic region, which is improbable in any real networks.
So, the behavior in the non-asymptotic regime is more important in practice.
Using ${\rm erfc}(x) \simeq \frac{1}{\sqrt{\pi}x} e^{-x^2}$ for
$x\gg 1$, we obtain that
\begin{equation}\label{phi_smallN}
\Phi_{SW} \simeq \frac{\alpha_1 \varepsilon \sqrt{\ln N}}{\alpha_2\sqrt{\pi}} 
\exp\left[ -\frac{\alpha_2^2}{\varepsilon^2\ln N}\right] 
\end{equation} 
for $N\ll N_c$. It has an essential singularity at $\varepsilon=0$ and
increases continuously and extremely slowly. 
Therefore, the spanning tree of finite SW networks undergoes an 
infinite-order transition at $\varepsilon=0$.

\begin{figure}
\includegraphics[width=\columnwidth]{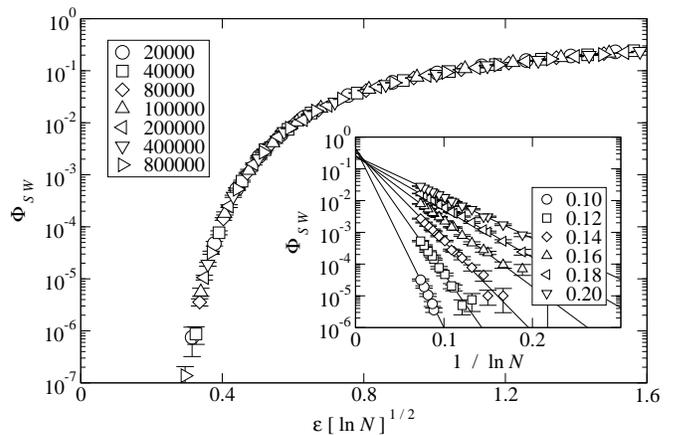}
\caption{Scaling plot of $\Phi_{SW}$ vs. $\varepsilon [\ln N]^{1/2}$ 
according to Eq.~(\ref{phi_sw}). 
The inset shows that $\ln \Phi \simeq a + b/\ln N$ 
for fixed values of $\varepsilon$, following Eq.~(\ref{phi_smallN}).
The straight lines are obtained from a least-square fitting.}
\label{fig:phi_sw}
\end{figure}

Numerical data are in good agreement with the mean-field results. 
In Fig.~\ref{fig:phi_sw}, we
plot $\Phi_{SW}$ as a function of the scaling variable 
$\varepsilon \sqrt{\ln N}$.
Data at different values of $N$ collapse onto a single curve, 
which supports the
scaling form in Eq.~(\ref{phi_sw}). In the inset, we plot $\Phi_{SW}$ at given
values of $\varepsilon$ against $1/\ln N$ for $N=100,\ldots,800000$ in the
semilog scale. 
We observe that the data align along straight lines, which support
the result in Eq.~(\ref{phi_smallN}). 
Extrapolating the straight lines to the limit $\ln N\rightarrow \infty$, 
we obtained that $\Phi_{SW}$'s converge to $ \alpha_1 \simeq 0.25$.

In the regular network, an activated edge may or may not increase the
separation of vertices from the source, which leads to the different scaling 
behaviors of $\Phi_{REG}\sim \varepsilon$ and $\Upsilon_{REG}\sim 
\varepsilon^2$. However, in the SW network, all activated edges considered 
above increase the separation. So $\Upsilon_{SW}$ shows the same scaling 
behavior as $\Phi_{SW}$, which we also confirmed numerically.

\section{Barab\'asi-Albert Scale-free network}\label{sec:BA}
In this section we study the response of the spanning tree of 
a SF network to the quenched disorder. 
We consider the Barab\'asi-Albert~(BA)
network with the degree-distribution exponent
$\gamma=3$~\cite{BA99}. 
Among all vertices, the {\em hub} which has the largest degree plays an
important role in the SF network~\cite{Goh01}. 
So we consider the spanning tree of the BA network 
with the hub as the source. $\Phi_{SF}$ is measured for the BA network 
with $Z_0=Z=4$, and compared with $\Phi_{REG}$ and 
$\Phi_{SW}$ in Fig.~\ref{fig:comp}. $\Phi_{SF}$ behaves
similarly to $\Phi_{SW}$.

Though the SF network is different from the SW network in many aspects, 
its spanning tree also has a random structure. 
Thus, we expect that the scaling
behavior of $\Phi_{SF}$ can be explained with the same mean-field-type
approximation as $\Phi_{SW}$. Under the approximation, each edge that does
not belong to ${\bf T}_s^0$ may be activated independently. 
Then, the mean density of the activated edges is obtained by replacing
$l(a,b)$ in Eq.~(\ref{p_ab}) with the diameter of the network.
Recently, it was found that the scale-free network is ultra small in
diameter~\cite{Cohen}. In particular, for $\gamma=3$ ($\gamma$: the 
degree-distribution exponent) the diameter scales as 
$D\sim \ln N/ \ln \ln N$, i.e., with a double logarithmic correction to 
the SW scaling $D\sim \ln N$. 
It leads us to conclude that 
\begin{equation}\label{phi_sf}
\Phi_{SF} = \alpha'_1\ {\rm erfc} \left(\frac{\alpha'_2}{\varepsilon \sqrt{
\ln N / \ln \ln N}}\right) 
\end{equation}
with constants $\alpha'_1$ and $\alpha'_2$.

\begin{figure}
\includegraphics[width=\columnwidth]{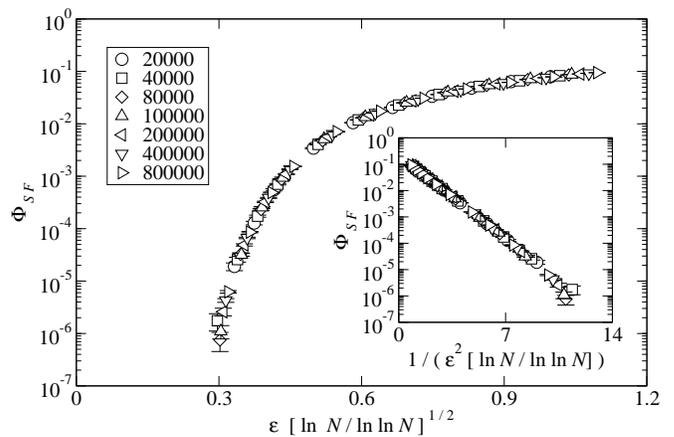}
\caption{Scaling plot of $\Phi_{SF}$ vs. $\varepsilon[\ln N/\ln\ln N]^{1/2}$
for the Barab\'asi-Albert networks with $Z=Z_0=2$
and $N=20000,\ldots,800000$. Inset shows a plot of $\Phi_{SF}$ vs.
$1/(\varepsilon^2 [\ln N/\ln\ln N])$ in semilog scale.}
\label{fig:phi_sf}
\end{figure}

This scaling form is indeed supported by our numerical data. 
In Fig.~\ref{fig:phi_sf}, we plot $\Phi_{SF}$ for the BA network with 
$Z=Z_0=2$ as a function of the scaling variable 
$\varepsilon \sqrt{\ln N/\ln\ln N}$, 
and obtain a good data collapse. Though the double-logarithmic correction is 
very weak, the data do not collapse at all without it. 
We note that the data also scale well with a scaling variable 
$\varepsilon (\ln N)^{1/4}$, however, we presume that this is accidental 
since the ratio between $(\ln N/\ln\ln N)$ and $(\ln N)^{1/2}$ remains 
almost constant up to $N=800000$.
The inset in Fig.~\ref{fig:phi_sf} shows that $\ln \Phi_{SF}$ is linear 
in $1/(\varepsilon^2 \ln N/\ln\ln N)$. 
It implies that the scaling function 
has the essential singularity as in Eq.~(\ref{phi_smallN}).
The same scaling behavior is observed universally for other
values of $Z=Z_0 = 3$ and $4$, and also for $\Upsilon_{SF}$.

The SF network shares the same scaling behavior with the SW network 
with the scaling variable $\varepsilon [\ln N/\ln\ln N]^{1/2}$.
The spanning tree of the SF network
undergoes a discontinuous transition in the asymptotic regime:
$\Phi_{SF}$ and $\Upsilon_{SF}$ in the infinite SF networks jump from zero 
at $\varepsilon=0$ to finite values as the disorder turns on. 
The asymptotic behavior sets in only when $N\gg N_c\sim \exp[\alpha
|\ln \varepsilon| \varepsilon^{-2}]$ with a positive constant $\alpha$.
In the transient regime with $N \ll N_c$, the spanning tree of the SF
networks undergoes a continuous infinite-order transition with the essential
singularity in $\Phi_{SF}$ and $\Upsilon_{SF}$ at $\varepsilon=0$.
Note that the crossover size $N_c$ of the SF network grows much faster than
that of the SW network. Therefore we conclude that the scale-free network
has the most stable transport pattern.

\section{Conclusion}\label{sec:conclusion}
In a network the shortest paths between vertices play an important role.
The optimal transport pattern from/to a vertex $s$ is
characterized by the spanning tree ${\bf T}_s$, a set of all shortest paths.
We have investigated the response of the spanning tree of complex networks 
to a quenched randomness in edge costs, and 
found that quenched disorder is a relevant
perturbation.
As the disorder turns on, the spanning tree evolves with a finite
fraction of edges modified and the network diameter expands.

For the regular network, the shape of the spanning tree can be described 
exactly using the random-walk mapping.
The spanning tree undergoes a continuous transition with
$\Phi_{REG} \sim \varepsilon$ and $\Upsilon_{REG} 
\sim \varepsilon^2$ with $\varepsilon$ the disorder strength.
For the SW and SF networks, we obtain the scaling form 
$\Phi(\varepsilon,N) = \widetilde{\Phi}(\varepsilon D^{1/2})$
where $D$ is the network diameter. It scales as $D \sim \ln N$
in the SW network and  $D\sim \ln N/\ln\ln N$ in the
BA network~\cite{Cohen}. 
The diameter-expansion coefficient satisfies the
same scaling form. The scaling function approaches a constant as its
argument $x=\varepsilon D^{1/2}\rightarrow \infty$. 
It has the essential singularity, $\widetilde{\Phi}(x) \sim
\exp[-\alpha x^{-2}]$ with a positive constant $\alpha$, as $x\rightarrow 0$.
Therefore, $\Phi$ and $\Upsilon$ have a {\em discontinuous} jump at 
$\varepsilon=0$ in infinite-size SW and SF networks. 
It shows that the SW network and the SF network are more affected by the 
quenched disorder than the regular network. 
However, the asymptotic scaling behavior emerges 
only when the network size is larger than the crossover size $N_c$.
It grows as $N_c \sim \exp[\alpha \varepsilon^{-2}]$ for the SW network 
and $N_c \sim \exp[ \alpha |\ln \varepsilon| \varepsilon^{-2}]$ for the
BA network with a positive constant $\alpha$ as
$\varepsilon$ vanishes. 
Since the crossover size grows extremely fast as $\varepsilon$ goes to
zero, the behavior in the transient regime with 
$N\ll N_c$ is more important in practice. 
In contrast to the discontinuous jump when $N\gg N_c$,
$\Phi$ at $N\ll N_c$ grows continuously and very slowly 
with the essential singularity at $\varepsilon=0$. Numerically we obtained 
that $\Phi < 10^{-7}$ for $\varepsilon \lesssim 0.1$ up to $N=800000$.
Therefore, we conclude that the transport pattern in 
finite SW and SF networks is extremely robust against 
the quenched disorder. In particular, the SF network is most stable.

The interesting scaling behavior of the SW and SF networks can be 
explained within a mean-field picture. $\Phi$ and $\Upsilon$ are 
proportional to the probability that 
a displacement of a random walker with diffusion constant 
${\cal O}(\varepsilon^2)$ is larger than 1 after $D$, the network diameter, 
time steps.
The crossover size is determined from the condition 
$\varepsilon^2 D(N_c) \simeq 1 $, and the spanning tree is very stable 
against the disorder as long as $N \ll N_c$.
So, the exponential divergence of the crossover size $N_c$ and hence
the extreme stability of the spanning tree are the
direct consequence of the small-worldness, i.e., $D(N) \sim \ln
N$~\cite{Newman00} and $D(N)\sim \ln N/\ln \ln N$~\cite{Cohen}, 
for the small-world and the scale-free network, respectively.
The diameter of the SF networks with $2<\gamma<3$ scales as $D(N) \sim
\ln\ln N$~\cite{Cohen} and within the mean-field picture one expects
the crossover size of these networks to scale like 
$N_c \sim \exp [ \exp [-\alpha \varepsilon^{-2}]]$.

Another consequence of the disorder is that the shortest path from one vertex
to the other becomes unique. 
Recently, it has been reported that 
the load distribution in SF networks follows a power-law
distribution with the universal load-distribution exponent $\eta\simeq 2.2$
or $2.0$~\cite{Goh01,Goh02}. 
Phenomenologically, the SF networks with degenerate shortest paths seem to
have $\eta \simeq 2.2$~\cite{Goh01,Goh02}, whereas those without or only 
a few degenerate shortest paths seem to have $\eta \simeq
2.0$~\cite{Goh02,Szabo02}. 
It would be interesting to study the load distribution in the disordered SF
networks, which will shed light on the role of the degeneracy on the
universality of the load distribution.
This work is in progress.

\acknowledgments
We would like to thank B. Kahng for useful discussions.
This work has been financially supported by the 
Deutsche Forschungsgemeinschaft (DFG).

\begin{appendix}
\section{Random walk in the presence of absorbing wall}
Consider a discrete-time random walk in one dimension 
in the presence of absorbing walls at $x=0$ and $x=L$. The walker is killed
upon a collision with the absorbing walls. Initially it is at
$x=a$ and its position after $t$ steps is given by 
$x(t) = a + \sum_{i=1}^t x_i$ with random
variables $x_i$'s with $\langle x_i\rangle=0$ and $\langle
x_i^2\rangle=\sigma^2$.
In the time-continuum limit, the probability density $P(x,t)$ satisfies a 
diffusion equation
${\partial P}/{\partial t} = \mu {\partial^2 P}/{\partial x^2}$
with a diffusion constant $\mu = {\sigma^2}/{2}$.
The absorbing walls impose boundary conditions $P(0,t)=P(L,t)=0$, 
and the initial condition implies $P(x,0) = \delta(x-a)$. 
In the Fourier series expansion
$P(x,t) = \sum_{n=1}^{\infty} p_n(t) \sin \frac{n\pi x}{L}$,
the diffusion equation becomes ${dp_n}/{dt} = -(\mu \pi^2 /L^2) n^2 p_n $
with the solution $p_n(t) = p_n(0) e^{-(\pi^2\mu/L^2) n^2 t}$. 
The coefficients $p_n(0)$ are determined from the initial
condition. Using the Fourier series expansion of the
delta function, we obtain that 
\begin{equation}\label{solution}
P(x,t) = \frac{2}{L} \sum_{n=1}^\infty \left(\sin\frac{n\pi a}{L}\right) 
\left(\sin \frac{n \pi x}{L}\right)e^{-\mu'n^2 t} \  . 
\end{equation}

The survival probability $S(t)$ of the random walker 
is given by the spatial integral of the probability density: 
$$S(t)  = \frac{4}{\pi}\sum_{n=0}^\infty
\frac{\sin((2n+1)\pi a/L)}{2n+1} e^{-(\pi^2 \mu/L^2)(2n+1)^2 t}\ .
$$
Then the death  probability during $t$ and $t+dt$
is given by $S(t)-S(t+dt) \simeq D(t) dt$ with 
$D(t) \equiv -\frac{dS}{dt}$. Therefore the mean life time of the 
walker is given as $\tau = \int_0^\infty dt~ t D(t) = 
\int_0^\infty dt S(t)$. 
Using (\ref{solution}), one obtains that
\begin{equation}\label{tau}
\tau = \frac{8L^2}{\sigma^2 \pi^3} \sum_{n=0}^\infty 
\frac{\sin((2n+1)\pi a/L)}{(2n+1)^3} \ .
\end{equation}

It becomes simpler in a special case. 
When $a = L/2$, $\tau = \frac{8L^2}{\sigma^2 \pi^3} \beta(3)$ with 
$\beta(s) \equiv \sum_{n=0}^\infty (-1)^n (2n+1)^{-s}$. 
Using $\beta(3) = \pi^3/32$~\cite{Abramowitz}, one obtains that
\begin{equation}\label{tau1}
\tau = \frac{L^2}{4\sigma^2} \ .
\end{equation}
When $a \ll L$, one can use an expansion $\sin(n\pi a/A) 
\sim n\pi a/A$, which yields that 
$\tau = \frac{8a L}{\sigma^2 \pi^2} \sum_{{\rm odd\ } n} n^{-2}$.
Using that $\sum_{{\rm odd\ }n}n^{-2} = \frac{3}{4}\zeta(2)=\frac{\pi^2}{8}$
with Riemann zeta function $\zeta(s)\equiv \sum_{n=1}^\infty
n^{-s}$~\cite{Abramowitz},
one obtains 
\begin{equation}\label{tau2}
\tau = \frac{aL}{\sigma^2}\ .
\end{equation}

\end{appendix}


\begin{references}
\bibitem{Strogatz01} S.H. Strogatz, Nature {\bf 410}, 268 (2001).
\bibitem{AB02} R. Albert and A.-L. Barab\'asi, 
        Rev. Mod. Phys. {\bf 74}, 47 (2002).
\bibitem{Watts98} D.J. Watts and S.H. Strogatz, 
        Nature {\bf 393}, 440 (1998).
\bibitem{BA99} A.-L. Barab\'asi and R. Albert, Science {\bf 286}, 509 (1999).
\bibitem{BAsolution} A.-L. Barab\'asi, R. Albert, and H. Jeong, 
           Physica A {\bf 272}, 173 (1999); 
        P.L. Krapivsky, S. Redner, and F. Leyvraz, 
           Phys. Rev. Lett.  {\bf 85}, 4629 (2000);
        S.N. Dorogovtsev, J.F.F. Mendes, and A.N. Samukhin, 
           Phys. Rev. Lett. {\bf 85}, 4633 (2000).
\bibitem{Dorogovtsev} S.N. Dorogovtsev, A.V. Goltsev, and J.F.F. Mendes, 
        cond-mat/0203227.
\bibitem{Igloi} F. Igl\'oi and L. Turban, cond-mat/0206522.
\bibitem{KimBJ} B.J. Kim, H. Hong, P. Holme, G.S. Jeon, P. Minnhagen, 
        and M.Y. Choi, Phys. Rev. E {\bf 64}, 056135~(2001).
\bibitem{Pastor-Satorras} R. Pastor-Satorras and A.  Vespignani, 
         Phys. Rev. Lett. {\bf 86}, 3200 (2001).
\bibitem{Albert00} R. Albert, H. Jeong, and A.-L. Barab\'asi, 
        Nature {\bf 406}, 378 (2000).
\bibitem{Cohen00} R. Cohen, K. Erez, D. ben-Avraham, and S. Havlin,
        Phys. Rev. Lett. {\bf 85}, 4626 (2000).
\bibitem{Callaway00} D.S. Callaway, M.E.J. Newman, S.H. Strogatz, 
        and D.J. Watts, Phys. Rev. Letts. {\bf 85}, 5468 (2000).
\bibitem{Goh01} K.-I. Goh, B. Kahng, and D. Kim, 
        Phys. Rev. Lett. {\bf 87}, 278701 (2001).
\bibitem{Szabo02} G. Szab\'o, M. Alava, and J. Kert\'esz, 
        Phys. Rev. E {\bf 66}, 026101 (2002).
\bibitem{Newman01} M.E.J. Newman, Phys. Rev. E {\bf 64}, 016131 (2001);
        {\it ibid.} {\bf 64}, 016132 (2001).
\bibitem{Goh02} K.-I. Goh, E.S. Oh, H. Jeong, B. Kahng, and D. Kim,
        cond-mat/0205232.
\bibitem{Heiko} M. Alava, P.M. Duxbury, C. Moukarzel, and H. Rieger,
        in {\it Phase Transitions and Critical Phenomena},
        edited by C. Domb and J.L. Lebowitz (Academic Press, Cambridge, 2000), 
        Vol. 18, p. 141-317; 
        A. Hartmann and H. Rieger, {\it Optimization algorithms in physics},
        (Wiley-VCH, Berlin, 2002).
\bibitem{Yook01} S.H. Yook, H. Jeong, A.-L. Barab\'asi, and Y. Tu, 
        Phys. Rev. Lett. {\bf 86}, 5835 (2001).
\bibitem{comment} The spanning tree has another ladder for 
vertices and $c_n$ and $d_n$. At small disorder, the two branches evolve
independently and do not overlap. 
So it suffices to consider only the single branch for $a_n$ and $b_n$.
\bibitem{Schroder} G. Schr\"oder, T. Knetter, M. Alava, and H. Rieger,
        Eur. Phys. J. B {\bf 24}, 101 (2001). 
\bibitem{Newman00} M.E.J. Newman, C. Moore, and D.J. Watts, 
        Phys. Rev. Lett. {\bf 84}, 3201 (2000).
\bibitem{Cohen} R. Cohen and S. Havlin, cond-mat/0205476.
\bibitem{Abramowitz} M. Abramowitz and I. A. Stegun, 
        {\it Handbook of Mathematical Functions} (Dover Publications Inc.,
        New York, 1970).

\end{references}
\end{document}